\documentclass[aip, preprint, groupedaddresses,superscriptaddress]{revtex4-1} 
\usepackage[utf8]{inputenc}

\usepackage{graphicx}% Include figure files
\usepackage{dcolumn}% Align table columns on decimal point
\usepackage{bm}% bold math
\usepackage{array}
\usepackage{multirow}
\begin{document}

\title{Improved initial guess for minimum energy path calculations}

\author{S{\o}ren Smidstrup}
\affiliation{QuantumWise A/S, Lers{\o} Parkall\'e 107, DK-2100 Copenhagen, Denmark}
\affiliation{Science Institute and Faculty of Physical Sciences, University of Iceland VR-III, 107 Reykjav\'{\i}k, Iceland.  \ \ e-mail: hj@hi.is}
\author{Andreas Pedersen}
\affiliation{Science Institute and Faculty of Physical Sciences, University of Iceland VR-III, 107 Reykjav\'{\i}k, Iceland.  \ \ e-mail: hj@hi.is}
\affiliation{Integrated Systems Laboratory, ETH Zurich, 8092 Zurich, Switzerland}
\author{Kurt Stokbro}
\affiliation{QuantumWise A/S, Lers{\o} Parkall\'e 107, DK-2100 Copenhagen, Denmark}
\author{Hannes J\'onsson}
%\email{hj@hi.is}
\affiliation{Science Institute and Faculty of Physical Sciences, University of Iceland VR-III, 107 Reykjav\'{\i}k, Iceland.  \ \ e-mail: hj@hi.is}
\affiliation{Dpt. of Applied Physics, Aalto University, FI-00076, Finland}

%\eads{\mailto{andreas@theochem.org}}

\begin{abstract}
A method is presented for generating a good initial guess of a transition path
between given initial and final states of a system without evaluation of the energy.
An objective function surface is constructed using an interpolation of pairwise distances at each discretization point along the path and the nudged elastic band method then used to find an optimal path on this
image dependent pair potential (IDPP) surface. This provides an initial path for the more computationally intensive
calculations of a minimum energy path on an energy surface obtained, for example, 
by {\it ab initio} or density functional theory.
The optimal path on the IDPP surface is significantly closer to a
minimum energy path than a linear interpolation of the Cartesian coordinates and, therefore, reduces the
number of iterations needed to reach convergence and averts divergence
in the electronic structure calculations when atoms are brought too close to each other in the initial path.
The method is illustrated with three examples:
(1) rotation of a methyl group in an ethane molecule,
(2) an exchange of atoms in an island on a crystal surface, and
(3) an exchange of two Si-atoms in amorphous silicon.
In all three cases, the computational effort in finding the minimum energy path with DFT was reduced by a factor
ranging from 50\% to an order
of magnitude by using an IDPP path as the initial path.
The time required for parallel computations was reduced even more
because of load imbalance when linear interpolation of Cartesian coordinates was used.
\end{abstract}

\maketitle
%---------------------------------------------------------------------------------------------------------------------

\section{Introduction}
\label{sec:Introduction}

Theoretical studies of the transition mechanism and rate of thermally activated events involving
rearrangements of atoms often involve finding the minimum energy path (MEP) for the system to advance from the
initial state minimum to the final state minimum on the energy surface. The estimation of the transition rate
within harmonic transition state theory requires finding the highest energy along the MEP, which
represents a first order saddle point on the energy surface.  While it is possible to use various methods to converge directly onto a saddle point starting from some initial guess,
methods that produce the whole minimum energy path are useful because one needs to ensure that the highest
saddle point for the full transition has been found, not just some saddle point.  Furthermore, calculations of MEPs
often reveal unexpected intermediate minima.

The nudged elastic band (NEB) method is frequently used to find MEPs \cite{NEB,Henkelman00a,Henkelman00b}.
There, the path is represented by a discrete
set of configurations of all the atoms in the system.
Some examples of complex MEPs calculated using NEB can, for example, be found in
refs.\cite{Batista01,Henkelman06,Jonsson11}.
The method is iterative and requires as input some initial path
which is then refined to bring the configurations to the MEP.
Harmonic spring forces are included between adjacent images to distribute them along the path. If all the spring constants
are chosen to be the same, the images will be equally distributed.  Other choices can be made, for example
higher density of images in the higher energy regions \cite{Henkelman00a}.
A force projection, the nudging, is used to remove the component of the true force parallel to the path and the perpendicular component of the spring force \cite{NEB}.
The number of iterations required can strongly depend on how close the initial path is to the MEP.  An important aspect of the method is that no knowledge of the transition mechanism is needed, only the endpoints corresponding to the initial and final states.
If more than one MEP exists between the given endpoints, the NEB will most likely converge on the one closest to the initial guess.  The initial path can, therefore, in some cases affect the results obtained with the NEB method.

The initial path in an NEB calculation has so far typically been constructed by a linear interpolation (LI) of the Cartesian
coordinates between the initial and final state minima.
The method is often used for systems subject to periodic boundary conditions were internal coordinates are not practical.  Such paths can, however, involve images where atoms come much too close
together, leading to large atomic forces and even divergence in electronic structure calculations.
One simple way of avoiding such problems is to check interatomic distances in each image and move atoms apart if the distance between them is shorter than a cutoff value.
Other methods for finding MEPs have also been devised where paths are constructed in a sequential manner, 
adding one image after another
with some type of extrapolation and relaxation each time a new image is added \cite{Peters04, Behn11}.

Here, we present a method for generating paths which are better suited as initial paths in NEB calculations than LI paths.
It is inspired by one of the earliest approaches for calculating reaction paths,
the one presented by Halgren and Lipscomb \cite{Halgren77}.
Their method involved two steps.  First, a linear synchronous transit (LST) pathway was constructed so as to make pair distances change gradually along the path (see below) and then an optimization
procedure, the quadratic synchronous transit, was carried out to further refine the path.  In the method presented here, the basic idea of LST is used to generate an improved initial guess for the NEB method, but the procedure is different
from the one used by Halgren and Lipscomb, as explained below.

The article is organized as follows:  In the following section, the LI and LST methods are reviewed and the new method  presented.  In section 
\ref{sec:applications},
%III, 
three applications are presented,
(1) rotation of a methyl group in an ethane molecule,
(2) an exchange of atoms in an island on a crystal surface, and
(3) an exchange of two Si-atoms in amorphous silicon.
The article concludes with a summary in section 
%IV.
\ref{sec:Conclusions}.

%---------------------------------------------------------------------------------------------------------------------

\section{Initial path generation}
\label{sec:methods}

The NEB method involves finding a discrete representation of the MEP.  First,  the
atomic coordinates at the two endpoints, i.e. the $3N$ coordinates of the $N$ atoms at the energy minima representing
initial and final states of the transition,
$r_\alpha$ and $r_\beta$,
are used to generate an initial path.
Here, $r$ will denote the vector of $3N$ coordinates of the atoms in a
given configuration, $r=\{r_1, r_2, \dots r_N\}$. Typically, a linear interpolation of the Cartesian coordinates of the two endpoint configurations is used as a starting guess.  Given that $p-1$ intermediate discretization points, here referred to as `images' of the system, will be used, the LI path which so far has been most commonly used as initial path in
NEB calculations, is given by
\begin{eqnarray}
\label{equ:linearinter}
r^\kappa_{L,i} = r^\alpha_i + \kappa  {{(r^\beta_i-r^\alpha_i)} / p}
\end{eqnarray}
Here, $r_i$ denotes the coordinates of atom $i$ and the index $\kappa$ denotes the image number in
the path and runs from $1$ to $p -1$.  In an NEB calculation, a minimization procedure is then carried out to adjust the coordinates of the $p-1$
intermediate images until they lie on the MEP, while the endpoint images are kept fixed. As mentioned above, there can be problems starting a calculation from an LI path, especially when electronic structure calculations are
used to evaluate the energy and atomic forces, since atoms can land too close to each other, leading to large atomic forces or even convergence problems in the electronic self-consistency iterations.
If two atoms are too close and need to be moved apart in an image, $\mu$, or
if one chooses to make use of some knowledge of a reasonable intermediate configuration, $r_{\mu}$, then the initial path for the NEB can be constructed
by first creating a linear interpolation from $r_\alpha$ to $r_{\mu}$ and then from $r_{\mu}$ to $r_\beta$.  But, it is better to have an automatic way, 
as presented in the following section, 
of creating a path where pair distances are automatically physically reasonable, 
and where the initial path is more likely to lie closer to the MEP than a linear interpolation,
thereby reducing the number of iterations needed to reach convergence.
%Before presenting the method, we first review the LST method of Halgren and Lipscomb which inspired the method
%presented here, and then present the new method.

% ----

\subsection{Image dependent pair potential}

Following the first step in the two step procedure presented by
Halgren and Lipscomb \cite{Halgren77},
% knowledge about the shape of the energy surface in the most relevant region.
an interpolation of all pair
distances between atoms is carried out for each of the intermediate images along the path.
These pair distances provide target values
which the initial path is then made to match as closely as possible.
The interpolated distance between atoms $i$ and $j$ in image $k$ is
\begin{eqnarray}
\label{equ:linearinter}
d^\kappa_{ij} = d^\alpha_{ij} + \kappa  {{(d^\beta_{ij}  -  d^\alpha_{ij})} / p}
\end{eqnarray}
where
$d_{ij}=\sqrt{\sum_\sigma (r_{i,\sigma}-r_{j,\sigma})^2}$ with $\sigma=x,y $ and $z$,
is the distance between atoms $i$ and $j$ in a given configuration of the atoms.
The LI path and the interpolation of pair distances are illustrated in Fig. 1.

%---------------------------------  Figure 1  ---------------------------
\begin{figure}
\centering
\includegraphics[width=.45\textwidth]{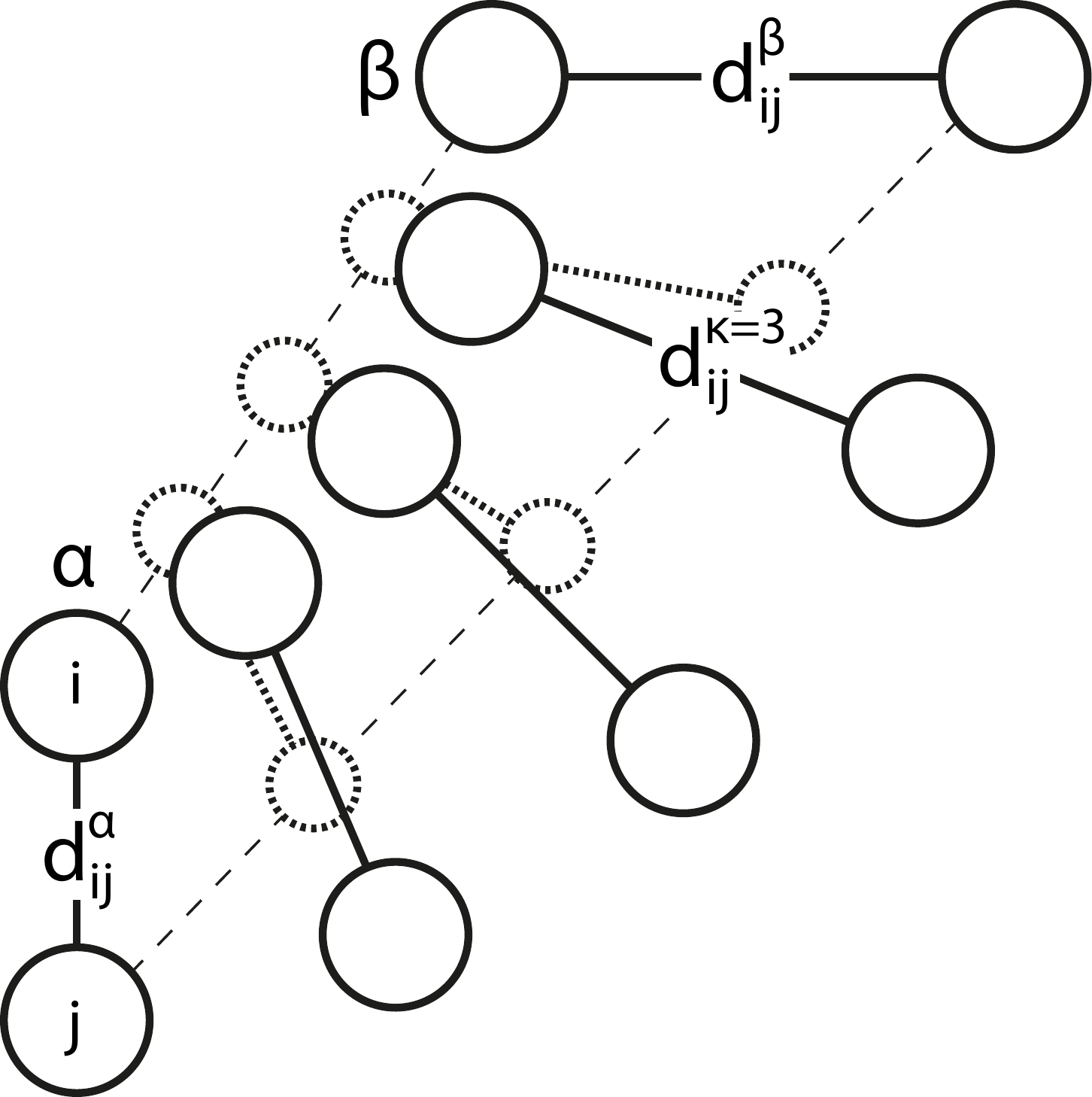}
\caption{
An illustration of paths generated by linear interpolation of Cartesian coordinates, LI (dashed),
and by interpolation of distances between atom pairs, IDPP (solid).
}
\label{fig:fig1}
\end{figure}
%----------------------------------------------------------------------------------------

Since there are many more atom pair distances than atomic degrees of freedom,
$N(N-1)/2$ vs. $3N-6$, the interpolated values of the atom coordinates
cannot satisfy the constraints rigorously and a compromise needs to be made.
An objective function can be defined for each image
by summing the squared deviation of pair distances from the target values
\begin{eqnarray}
\label{equ:IDPP}
S^{IDPP}_\kappa(r)  = \sum_i^N \sum_{j>i}^N \ w(d_{ij})
\left(d^\kappa_{ij} -\sqrt{ \sum_\sigma (r_{i,\sigma}-r_{j,\sigma})^2}\right)^2
\end{eqnarray}
%
%  / (r_{i}-r_{j})^4
Here, $w$ is a weight function which can be used to place more emphasis on short distances, since the
energy of an atomic system rises strongly when two atoms come too close together.
The function $S^{IDPP}$ defines an objective function for each image which has the form of a
pairwise interaction potential that directs the atom coordinates to a configuration where the
distances between atoms are close to the interpolated distances.
One can think of $S^{IDPP}_{\kappa}(r)$, as defining an effective
`energy surface' and use the NEB method to find the optimal path on the $S^{IDPP}_\kappa(r)$ surface.
The force acting on atom $i$ in image $\kappa$ is then obtained as
\begin{eqnarray}
\label{equ:IDPPforce}
F^\kappa_i (r) = - \nabla_i S^{IDPP}_{\kappa}(r)
\end{eqnarray}
After applying the NEB iterative minimization \cite{NEB} with all spring constants chosen to be equal, a path with even distribution of the images is obtained where atom pair distances are changing gradually from one image to another.
We will refer to this as the IDPP path.
As we illustrate below, with three example calculations, the IDPP path is closer to the MEP on the true energy surface than the LI path. By using the IDPP path as an initial path for NEB calculations using atomic forces obtained by
density functional theory (DFT), the number of iterations needed to reach convergence was, in the cases studied here, significantly smaller than when the LI path was used as the initial path.
One good aspect of the IDPP method is that it
does not require a special coordinate system, such as internal coordinates (e.g., bond distances and angles) and, therefore, has the advantage of being easily applicable to any system, including systems subject to periodic boundary
conditions.

%  ------------

\subsection{LST path}

We now compare this procedure to what
Halgren and Lipscomb named the LST path \cite{LSTdisc}.  There, an objective function
was defined as
\begin{eqnarray}
\label{equ:LST}
S^{LST}_\kappa (r) = S^{IDPP}_{\kappa}(r) +  \gamma \sum_i^N \ (\sum_\sigma r_{i,\sigma}-r^\kappa_{L,i,\sigma})^2
\end{eqnarray}
with the parameter $\gamma$ chosen to be $10^{-6}$ in atomic units \cite{Halgren77}.
The weight function in $S^{IDPP}$ was chosen to be
$w(d)=1/d^4$.
%
%\begin{eqnarray}
%\label{equ:weight}
%w(r_i,r_j)  = {1 \over { \sum_\sigma (r_{i,\sigma}-r_{j,\sigma})^2} }
%\end{eqnarray}
%
This choice for the function places greater weight on short distances, which are more important
for the energetics.
We have chosen the same weight function here in our calculations of the IDPP paths.
The second term on the right hand side of eqn.(\ref{equ:LST}) was added to remove uniform translation and
help make the path continuous.
(The $r^\kappa_{L,i}$ are given by eqn.(\ref{equ:linearinter})).
It is, however, quite arbitrary and does not necessarily result in a continuous path with an even distribution of the images, as illustrated below.
The atom coordinates, $r$, in each image, $\kappa$, were chosen so as to minimize $S^{LST}$, in a least squares
procedure \cite{Halgren77}.

%------------------------------

\section{Applications}
\label{sec:applications}

For illustration purposes, the method was applied to transitions in three different systems:
Rotation of a methyl group in ethane, interchange of atoms in a heptamer island on a surface and 
interchange of atoms in amorphous silicon.
In all cases the initial path was generated using the ATK software \cite{ATK}.
Once the initial path had been constructed, the NEB method was applied to find the MEP 
\cite{NEB,Henkelman00a,Henkelman00b}. 
The iterative NEB minimizations using velocity projections
(`quick-min') \cite{NEB,Sheppard08}
were carried out until the maximum force on each atom in any of the images had dropped below $0.5$~eV/\AA\
and then the climbing-image NEB was used with a conjugate gradient minimization
algorithm \cite{Henkelman00a,Sheppard08} until the maximum force dropped
below $0.1$~eV/\AA.
A tolerance of $0.1$~eV/\AA\ is typically sufficient to get a good estimate of the path. The calculations of the condensed
phase systems were carried out using
VASP \cite{VASP}, PBE functional \cite{PBE}  and PAW \cite{PAW}, with the TST tools \cite{TSTtools}.
The heptamer island system consisted of a slab of 3 layers, each with 36 atoms and the calculation included 
(3x3x1) k-points and an energy cutoff of 270 eV.  The calculation of the amorphous Si involved 214 atoms 
including only the gamma point and an energy cutoff of 245 eV.
The methyl rotation was calculated using the ATK-DFT software \cite{ATK}, PBE functional and linear combination of atomic orbitals (LCAO).

%  ----------------------------------------------------------------------------------------------------------------------------------------

\subsection{Rotation of a methyl group in ethane}

The first application is rotation of a methyl group in an ethane molecule. This simple example illustrates well the
difference between LI and IDPP paths, which are shown in Fig. 2 with 5 intermediate images. The constraint on the pairwise distances
in the construction using the IDPP path leads to a simple rotation of the methyl group,
while the LI method gives a path with significant variations in the C-H bond lengths. 
The convergence in a subsequent NEB calculation starting from
the IDPP path and using atomic
forces from an ATK-DFT\cite{ATK} calculation required about a third as many atomic iterations and SCF iterations to
reach convergence as compared with a calculation starting with the LI path, see Table 1.

%---------------------------------  Figure 2  ---------------------------
\begin{figure}
\centering
\includegraphics[width=.65\textwidth]{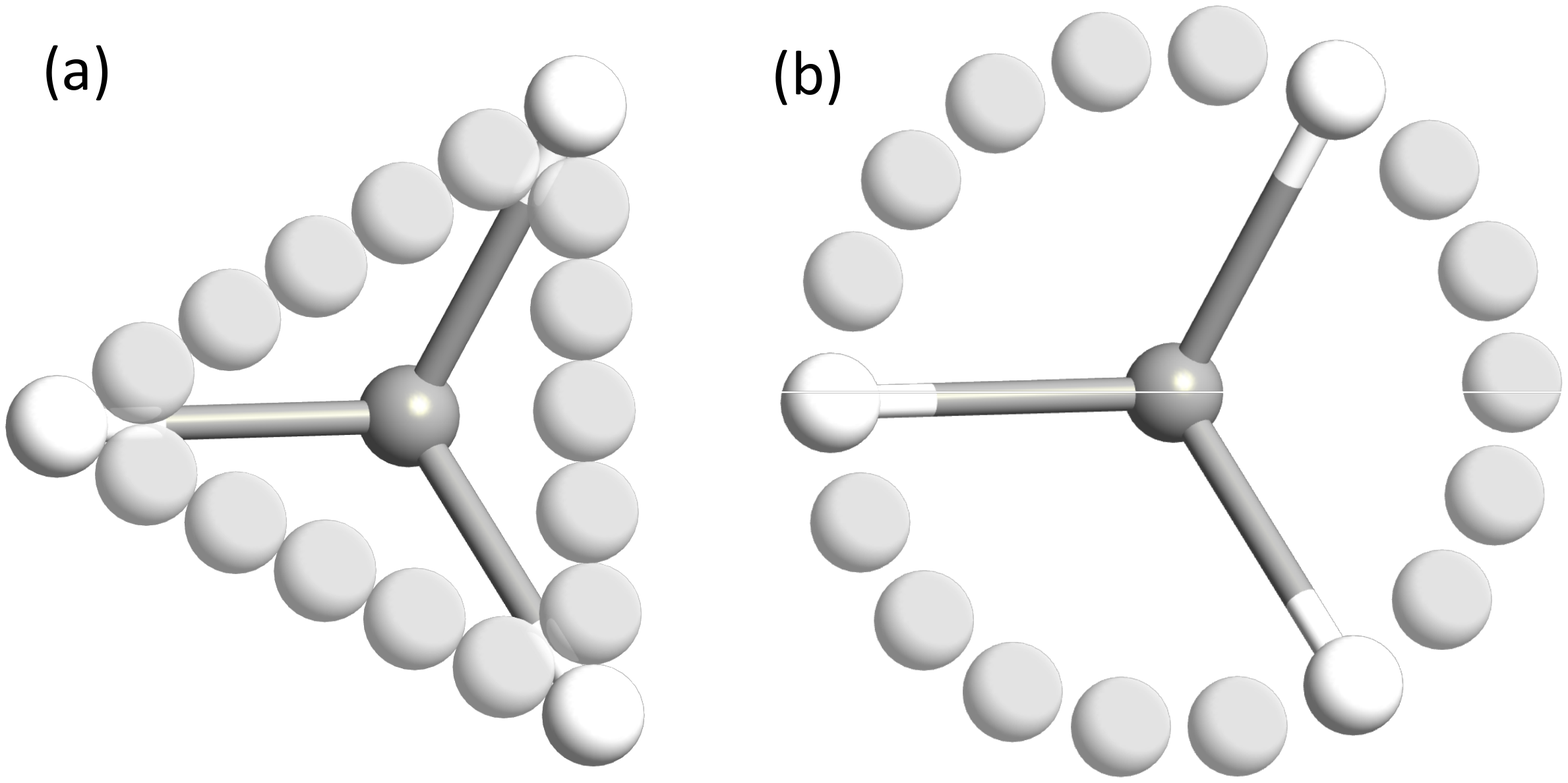}
\caption{
Initial path for the rotation of a methyl group in ethane,  \ (a) created by LI, a linear interpolation of Cartesian coordinates;
\ (b) created using the IDPP method. The rotation is clearly better represented by the latter method. The NEB calculation
starting from the IDPP path required about a third as many atomic iterations and SCF iterations to
reach convergence as compared with a calculation starting with the LI path, see Table 1.
}
\label{fig:fig2}
\end{figure}
%----------------------------------------------------------------------------------------

%  ------------------------------------------------------------------------------------------------------------------------------------------------

\subsection{Heptamer island on a surface}

The second application is an interchange of atoms in a heptamer island of six Al-atoms and one Ni-atom sitting
on an Al(111) surface. The transition involves the concerted movement of the Ni-atom from a rim site to the central
site of the island while the Al-atom initially in the center takes the rim site.  An LI path with an
odd number of images would have direct overlap of the two atoms in the image at the middle of the path
resulting in divergence of the subsequent DFT calculation.
In order to avoid that, the Ni-atom was moved out along the surface normal by 0.2 {\AA}
in both the initial and final state as 7 intermediate images in the LI path were generated, see Fig. 3(a).
The NEB calculation was then started from
a path where the Ni-atom was placed in the right position in the initial and final states but intermediate images
were taken from generated the LI path. 
Despite these adjustments, the NEB calculation required a large number of iterations, see Table 1.

A path constructed using the LST method of Hallgren and Lipscomb was also created and is shown in Fig. 3(b).
A discontinuity in the path is
evident from the figure because the constraint given by the second term in eqn.(\ref{equ:LST}) is too weak.
The IDPP path is shown in Fig. 3(c).  It has an even distribution of the images along the path, which otherwise is similar
to the LST path.  The final position of the images after NEB relaxation using forces obtained by DFT is shown in Fig. 3(d).
The displacements of the exchanging atoms are substantially larger than in the IDPP path, and
neighboring atoms in the island turn out to undergo large displacement in the intermediate images,
which is missing in the IDPP because
their position is nearly the same in the initial and final state.  The NEB/DFT calculation started from the IDPP path
required half as many iterations as the calculation
started from the LI path.   More importantly, the number of electronic iterations, which is proportional to the CPU time,
was an order of magnitude smaller when the NEB/DFT calculation is carried out starting from the IDPP path, see Table 1.

The energy along the minimum energy path is shown in Fig. 4.  Interestingly, an intermediate minimum is identified by the
NEB calculation where the Ni-atom has moved on-top of the cluster while the Al-atom has pushed two of the rim atoms
away from the center of the island.  Such intermediate minima are often found in NEB calculations and the resulting MEP then has more than one maximum.  This illustrates the importance of calculating a full MEP for the transition rather than
just finding a first order saddle point.  If a saddle point search is carried out starting from the $\beta$ state, the lower
saddle point will most likely be found and the activation energy underestimated unless a further exploration of the MEP is carried out.

%---------------------------------  Figure 3  ---------------------------
\begin{figure}
\centering
\includegraphics[width=0.7\textwidth]{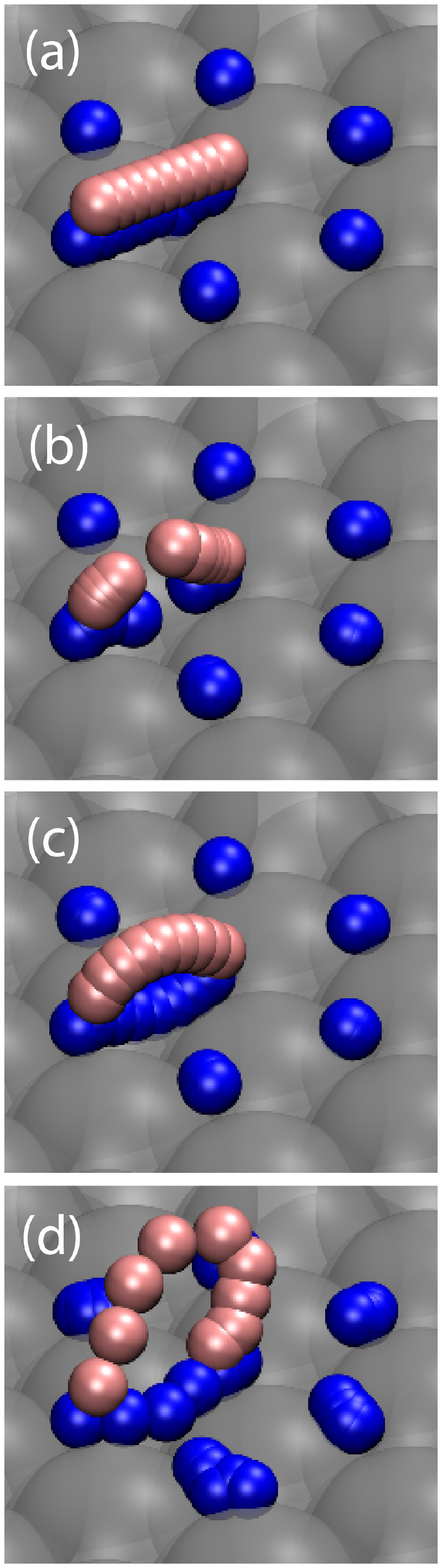}
\caption{
Paths for concerted rearrangement
of a Ni- and an Al-atom in a Al$_6$Ni heptamer island on an Al(111) surface.
(a) LI path created by linear interpolation of Cartesian coordinates of the initial and final states.  Here, the Ni-atom as been displaced outwards along the surface normal by 0.2~\AA\ in both the initial and final states to avoid a direct overlap
of atoms in the middle image and subsequent divergence of the DFT method.
(b) LST path generated using the method of Halgren and Lipscomb \cite{Halgren77}.
While the path avoids the direct overlap of atoms, the path is discontinuous with a large gap between two of the
intermediate images.
(c) IDPP path generated by an NEB calculation on the object function surface given by eqn. \ref{equ:IDPP}.
The path is continuous and has an equal spacing of the images.
(d) The minimum energy path found after NEB relaxation starting from any of the three paths shown in (a-c)
using atomic forces obtained from DFT.
}
\label{fig:fig3}
\end{figure}
%------------------------------

%---------------------------------  Figure 4  ---------------------------
\begin{figure}
\centering
\includegraphics[width=.45\textwidth]{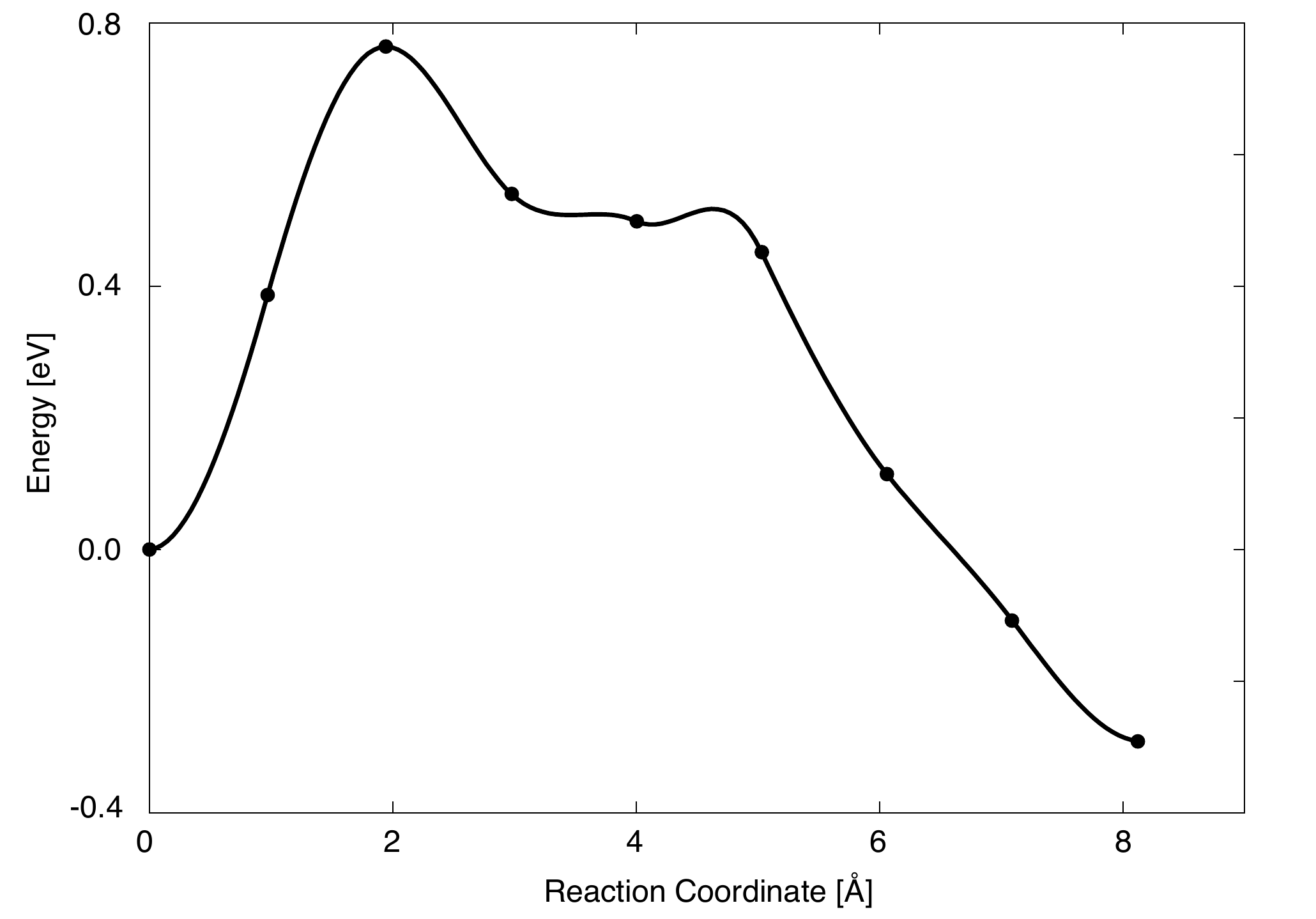}
\caption{
Energy along the minimum energy path for the concerted rearrangement
of a Ni- and an Al-atom in a Al$_6$Ni heptamer island on an Al(111) surface, obtained using the NEB method and DFT calculations of the atomic forces.
}
\label{fig:fig4}
\end{figure}
%----------------------------------------------------------------------------------------

%------------------------------ Table 1

\begin{table}
\caption{
Computational effort in converging to the minimum energy path starting from different initial paths:
the linear interpolation of Cartesian coordinates (LI)
and the Image Dependent Pair Potential (IDPP) method presented here.
The effort is reported as the number of electronic (SCF) iterations needed as well as the number of atomic displacement (AD) iterations.
In example 2, the Ni-adatom was moved outwards along the surface normal by
0.2~\AA\  in the intermediate images of the LI path to prevent the DFT calculation from diverging.
%The calculation was stopped when the magnitude of the atomic forces had dropped below 0.5~eV/{\AA}.
}
\vskip 0.4 true cm
\begin{tabular}{l | c c c c c c }
\hline
\hline
  	Application		& LI &&  IDPP  \\
\hline
Ex.1: Methyl rotation \\
SCF iterations	&  1802 &&     598   \\
AD iterations 	& 72      &&     28  \\
\hline
Ex.2: Heptamer island \\
SCF iterations	&  11367 &&  1120   \\
AD iterations 	& 151      && 71  \\
\hline
Ex.3: Si-atom exchange \\
SCF iterations	&  9413 &&         6433   \\
AD iterations 	& 149    &&          104  \\
\hline
\hline
\label{tab:barriers}
\end{tabular}
\end{table}
%
% ----------------------------------------------------------
% --------------------------------------------------------------------------------------------------------------------

\subsection{Exchange of Si-atoms in amorphous silicon}

In a third example, a calculation was carried out of a transition where two
Si-atoms in amorphous silicon change places in a
concerted way.  The sample which contained 214 Si-atoms was created by cooling a liquid and then
annealing \cite{PedersenInPrep}.
The path was calculated using a large number of intermediate images, p-1=15, to obtain a good resolution of this
rather complex path.
Since the two Si-atoms are changing places, the linear interpolation results in a small distance between the two atoms
in the middle image.
As a result, the energy obtained in the DFT calculations in the first few atom iterations is large and requires many
electronic iterations.
Since number of images in the path is large, the average number of electronic iterations
is only 50\% larger when staring form the LI path as compared to the calculation starting from the IDPP path, see Table 1.
But, the time required for the parallel NEB calculation is even longer because it is held up by this one, troublesome
image (Figs. 5 and 6).

%---------------------------------  Figure 5  ---------------------------
\begin{figure}
\centering
\includegraphics[width=.35\textwidth]{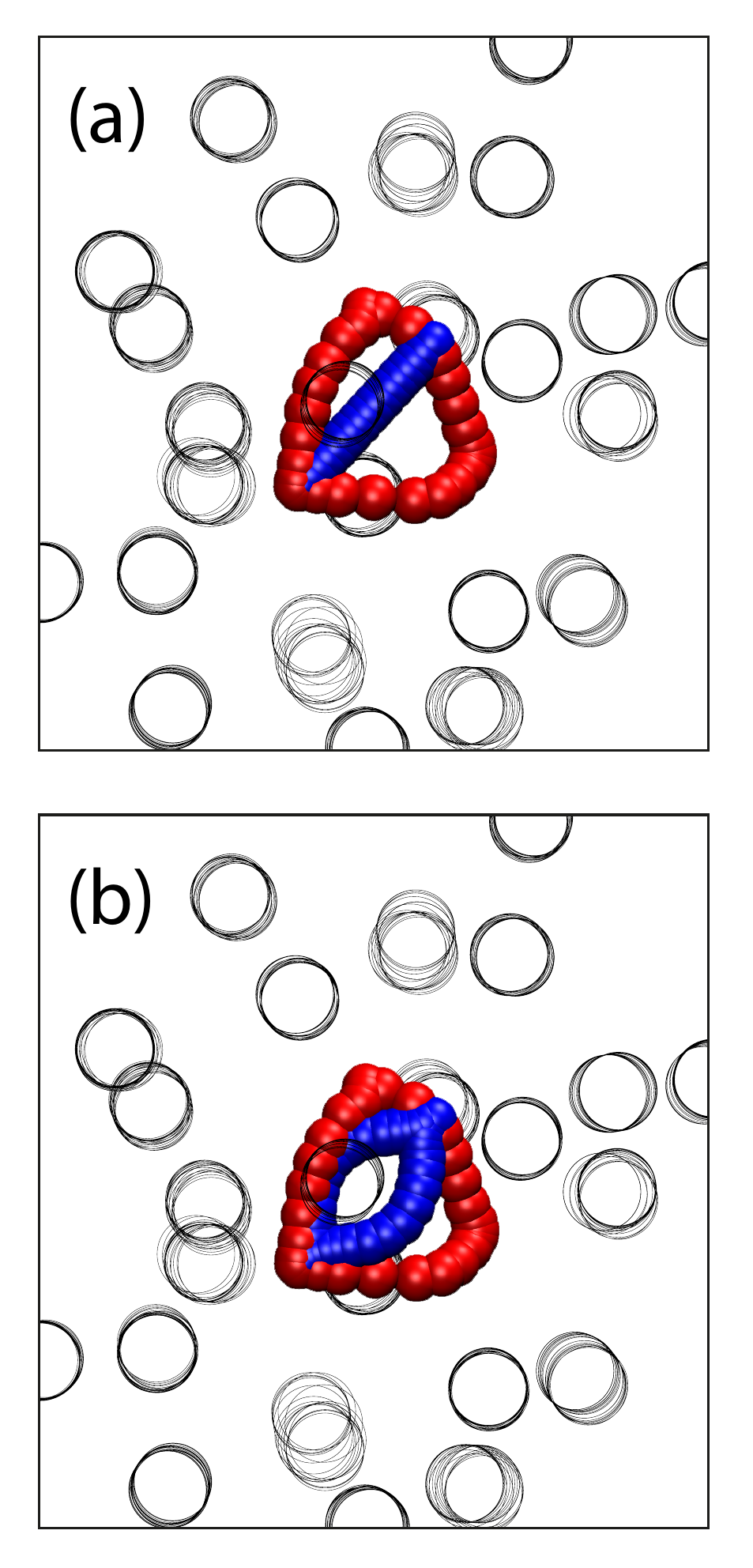}
\caption{
A diffusion event in amorphous silicon where two Si-atoms exchange places. (a) The LI path (blue).
(b) The IDPP path (blue).
The minimum energy path found by NEB method is also shown (red for the two Si-atoms that exchange places, circles for the neighboring Si-atoms).
}
\label{fig:fig5}
\end{figure}
%----------------------------------------------------------------------------------------

%---------------------------------  Figure 6  ---------------------------
\begin{figure}
\centering
\includegraphics[width=.45\textwidth]{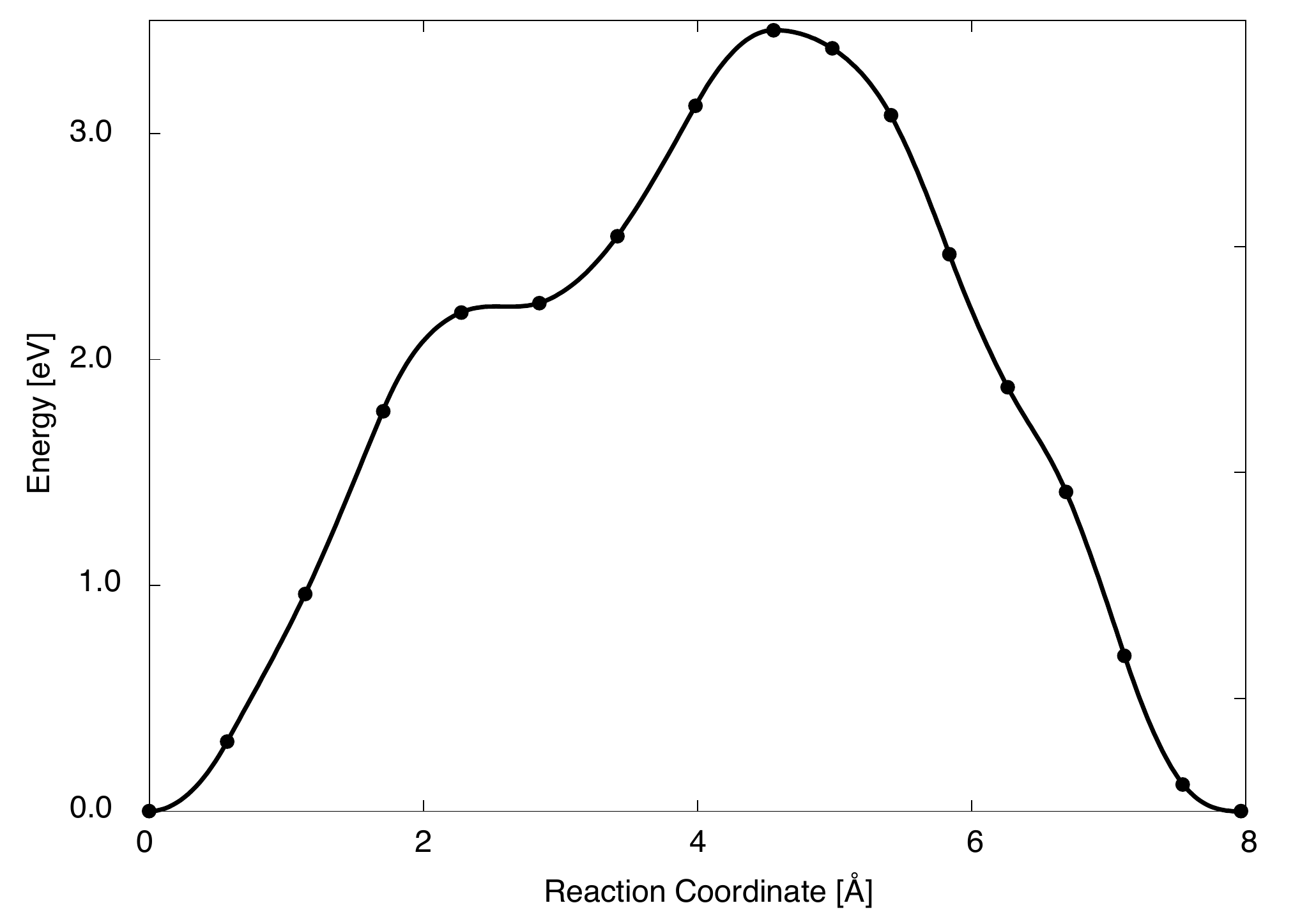}
\caption{
Energy along the minimum energy path for an exchange of two Si-atoms in amorphous silicon, shown in Fig. 5.
}
\label{fig:fig6}
\end{figure}
%--------------------------------------------------------------------------

% --------------------------------------------------------------------------------------------------------------------

\section{Conclusions}
\label{sec:Conclusions}

The IDPP method presented here is a robust and simple method for generating the input for NEB calculations
of minimum energy paths.  It can save considerable amount of computer time
as compared with
a linear interpolation between initial and final states using Cartesian coordinates. 
It can, furthermore, help avoid divergence in electronic structure calculations which can occur when the linear interpolation brings atoms too close together in an intermediate image of the initial path.  
The method is similar to the LST method of
Hallgren and Lipscomb, but is guaranteed to produce a continuous path as it makes use of the NEB method
on the objective function surface generated by interpolation of pair distances.  The method could
also be used to generate input for other path calculations, such as free energy paths where sampling of system 
configurations in hyperplanes is carried out either classically or quantum mechanically \cite{RW-QTST} and
for transitions between magnetic states where the orientation of magnetic moments changes \cite{spintransitions}.

Further development of the method could involve testing of other choices for the weight function, which was
simply taken here to be the one used by Hallgren and Lipscomb.  Also, testing in combination with other optimization 
methods in the NEB calculation, such as BFGS, would be useful.  
Since NEB calculations using {\it ab initio} or DFT evaluation of the atomic forces are 
computationally intensive, exploration of these and other ways of reducing the computational effort would be worthwhile.

% -------------
%\vskip -0.5 true cm

\section{Acknowledgements}

%\vskip -0.5 true cm

This work was supported by QuantumWise A/S, Eurostars project E6935 (ATOMMODEL) and the Icelandic Research Fund.
The calculations of the 214 Si-atom system were carried out using the Nordic High Performance Computing (NHPC) facility in Iceland. 

%---------------------------------------------------------------------------------------------------------------------
\newpage %Just because of unusual number of tables stacked at end


\begin{thebibliography}{25}


\bibitem{NEB}  H. J\'onsson, G. Mills and K. W. Jacobsen,
{\it Classical and Quantum Dynamics in Condensed Phase Simulations},
edited by B. J. Berne, G. Ciccotti, and D. F. Coker (World Scientific, Singapore, 1998), p. 385.

\bibitem{Henkelman00b} G. Henkelman and H. J\'onsson,
%  improved tangent
{\it J. Chem. Phys.} {\bf 113}, 9978 (2000).

\bibitem{Henkelman00a} G. Henkelman, B. P. Uberuaga, and H. J\'onsson,
%  CI-NEB
{\it J. Chem. Phys.} {\bf 113}, 9901 (2000).

\bibitem{Batista01}  E.R. Batista and and H. J\'onsson,
% Diffusion and Island formation on the ice Ih basal plane surface
{\it Comp. Mat. Sci.} {\bf 20}, 325  (2001).

\bibitem{Henkelman06}  G. Henkelman, A. Arnaldsson  and H. J\'onsson,
%  Theoretical calculations of CH4 and H2 associative desorption from Ni111:Could subsurface hydrogen play an important role?
{\it J. Chem. Phys.} {\bf 124}, 044706 (2006).

\bibitem{Jonsson11} H. J\'onsson,
{\it Proceedings of the National Academy of Sciences} {\bf 108}, 944 (2011).

\bibitem{Peters04}  B. Peters, A. Heyden, A. T. Bell and A. Chakraborty,
{\it J. Chem. Phys.} {\bf 120}, 7877 (2004).

\bibitem{Behn11}  A. Behn, P. M. Zimmerman, A. T. Bell and M. Head-Gordon,
{\it J. Chem. Theory Comput.} {\bf 7},  4019 (2011).
%  Incorporating Linear Synchronous Transit Interpolation into the Growing String Method: Algorithm and Applications

\bibitem{Halgren77} T. A. Halgren and W. N. Lipscomb,
{\it Chem. Phys. Lett.}, {\bf 49}, 225 (1977).
%THE SYNCHRONOUS-TRANSIT METHOD FOR DETERMINING REACTION PATHWAYS AND LOCATING MOLECULAR TRANSITION STATES

\bibitem{LSTdisc} We note that the phrase `linear synchronous transit' has later been used to refer to paths
constructed in quite different ways.

\bibitem{ATK}  ATK-DFT software, version 12.2.  See http://www.quantumwise.com

\bibitem{Sheppard08}  D. Sheppard, R. Terrell and G. Henkelman,
%  Optimization methods for finding minimum energy paths
%     conclude that BFGS is best.
{\it J. Chem. Phys.} {\bf 128}, 134106 (2008).

\bibitem{VASP} G. Kresse and J. Hafner,
{\it Phys. Rev. B}, \textbf{47}, 558 (1993);
G. Kresse and J. Furtm\"uller,
{\it Comput. Mater. Sci.}, \textbf{6}, 15 (1996).

\bibitem{PBE} J. P. Perdew and K. Burke and M. Ernzerhof,
{\it Phys. Rev. Lett.} {\bf 77}, 3865 (1996).

\bibitem{PAW} P. E. Bl\"ochl, 
{\it Phys. Rev. B} {\bf 50}, 17953 (1994).

\bibitem{TSTtools} http://theory.cm.utexas.edu/vasp/

\bibitem{PedersenInPrep}  A. Pedersen, L. Pizzagalli and H. J\'onsson,
(in preparation).

\bibitem{RW-QTST} G. K. Schenter, G. Mills, and H. J\'onsson,
%`Reversible Work Based Quantum Transition State Theory', 
{\it J. Chem. Phys.} {\bf 101},  8964 (1994);
G. Mills, G. K. Schenter, D. Makarov and H. J\'onsson, 
%`Generalized Path Integral Based Quantum Transition State Theory', 
{\it Chem. Phys. Letters} {\bf  278}, 91 (1997);
G. H. J\'ohannesson and H. J\'onsson, 
%`Optimization of Hyperplanar Transition States', 
{\it J. Chem. Phys.} {\bf 115}, 9644 (2001).

\bibitem{spintransitions} P. F. Bessarab, V. M. Uzdin and H. J\'onsson, 
{\it Zeitschrift f\"ur Physikalische Chemie} {\bf 227}, 1543 (2013);
{\it Phys. Rev. Letters} {\bf 110}, 020604 (2013);
{\it Phys. Rev. B. } {\bf 85}, 184409 (2012).


\end{thebibliography}
\end{document}